\documentclass[print,showpacs,oneside,twocolumn,superscriptaddress,prl,amsmath,amssymb]{revtex4}
\usepackage{cases}
\usepackage{amsmath}
\usepackage{amssymb}
\usepackage{amsfonts}
\usepackage{amssymb}
\usepackage{dcolumn}
\usepackage{bm}
\usepackage{hyperref}
\usepackage{graphicx}

\begin{document}
\title{Room-temperature implementation of the Deutsch-Jozsa algorithm with a single electronic spin in diamond}
\author{Fazhan Shi}
\affiliation{Hefei National Laboratory for Physics Sciences at
Microscale and Department of Modern Physics, University of Science
and Technology of China, Hefei, 230026, China}
\author{Xing Rong}
\affiliation{Hefei National Laboratory for Physics Sciences at
Microscale and Department of Modern Physics, University of Science
and Technology of China, Hefei, 230026, China}
\author{Nanyang Xu}
\affiliation{Hefei National Laboratory for Physics Sciences at
Microscale and Department of Modern Physics, University of Science
and Technology of China, Hefei, 230026, China}
\author{Ya Wang}
\affiliation{Hefei National Laboratory for Physics Sciences at
Microscale and Department of Modern Physics, University of Science
and Technology of China, Hefei, 230026, China}
\author{Jie Wu}
\affiliation{Hefei National Laboratory for Physics Sciences at
Microscale and Department of Modern Physics, University of Science
and Technology of China, Hefei, 230026, China}
\author{Bo Chong}
\affiliation{Hefei National Laboratory for Physics Sciences at
Microscale and Department of Modern Physics, University of Science
and Technology of China, Hefei, 230026, China}
\author{Xinhua Peng}
\affiliation{Hefei National Laboratory for Physics Sciences at
Microscale and Department of Modern Physics, University of Science
and Technology of China, Hefei, 230026, China}
\author{Juliane Kniepert}
\affiliation{Institut f{\"u}r Experimentalphysik, Freie
Universit{\"a}t Berlin, Arnimallee 14, 14195 Berlin, Germany}
\author{Rolf-Simon Schoenfeld}
\affiliation{Institut f{\"u}r Experimentalphysik, Freie
Universit{\"a}t Berlin, Arnimallee 14, 14195 Berlin, Germany}
\author{Wolfgang Harneit}
\affiliation{Institut f{\"u}r Experimentalphysik, Freie
Universit{\"a}t Berlin, Arnimallee 14, 14195 Berlin, Germany}
\author{Mang Feng}
\affiliation{State Key Laboratory of Magnetic Resonance and Atomic
and Molecular Physics, Wuhan Institute of Physics and Mathematics,
Chinese Academy of Sciences, Wuhan 430071, China}
\author{Jiangfeng Du}
\altaffiliation{djf@ustc.edu.cn} \affiliation{Hefei National
Laboratory for Physics Sciences at Microscale and Department of
Modern Physics, University of Science and Technology of China,
Hefei, 230026, China}
\begin{abstract}

The nitrogen-vacancy defect center (NV center) is a promising candidate for quantum information processing due to the possibility of coherent manipulation of individual spins in the absence of the cryogenic requirement. We report a room-temperature implementation of the
Deutsch-Jozsa algorithm by encoding both a qubit and an auxiliary state in the electron spin of a single NV center. By thus exploiting the specific $S=1$ character of the spin system, we demonstrate how even scarce quantum resources can be used for test-bed experiments on the way towards a large-scale quantum computing architecture.

\end{abstract}


\maketitle

Quantum computing (QC) outperforms its classical counterpart by exploiting quantum phenomena, such as superposition of states, entanglement and so on. Although the rudiments of QC are clear and some quantum algorithms have been proposed so far, implementation of QC is still experimentally challenging due to the decoherence induced by coupling to the environment. To avoid or suppress the decoherence, operations on most QC candidate systems that are considered scalable to a large number of qubits have been carried out at low temperatures. Despite that technical effort, only a few quantum gate operations could be achieved coherently within a single implementation.

Compared to other QC candidate systems, however, the nitrogen-vacancy defect center (NV center) in diamond is an exception where QC operations on individual spins could be achieved at room temperature \cite {nv}. Since the first report for optically detected magnetic resonance on single NV centers in 1997 \cite{first}, much progress has been achieved in exactly manipulating this system. As both electronic and nuclear spins are now well
controllable \cite {jele1,jele2,jele3,jele4,giga}, NV centers could be used as very good building blocks for a large-scale QC architecture. In the QC implementation, the electronic spins are manipulated in an optical fashion, and the nuclear spins are operated by means of hyperfine coupling. Currently available techniques have achieved the quantum information storage and retrieval between electron spin and the nuclear spins \cite{register}. This technique also enables rapid, high-fidelity readout of quantum information from the electron spin \cite{readout}. On the other hand, using the nuclear spins and additional electron spins as a controllable environment, a 'surprisingly different behavior' in the dynamics of the single electron spin was observed in different situations \cite{bath}. As a result, NV centers are considered as an excellent test bed for models and schemes of QC.

Despite all this progress in quantum gate realization, no real quantum algorithms have yet been demonstrated. In this Letter, we report a room-temperature implementation of a quantum algorithm, i.e., the refined Deutsch-Jozsa (RDJ) algorithm \cite {sdj}, using a single NV center. The RDJ algorithm is the simplified version of the original DJ algorithm \cite {dj}, one of the most frequently mentioned quantum algorithms. As the first proposed quantum algorithm, the DJ algorithm has been employed in different systems to demonstrate the exponential speed-up in distinguishing constant from balanced functions with respect to the corresponding classical algorithm. For example, it has been carried out experimentally in nuclear magnetic resonance systems \cite{nmr}, in quantum dot systems \cite {qd1,qd2}, by linear optics \cite {os}, and by trapped ions \cite {ion}. Compared to the original DJ algorithm, the refined version \cite {sdj} removes the qubit for the evaluation of the function, which remains unchanged during the algorithm implementation. As a result, it can reduce the required qubit resources, but still maintain the superiority due to quantum power over the corresponding classical means.

\begin{figure}[htbp]
\centering
\includegraphics[width=1\columnwidth]{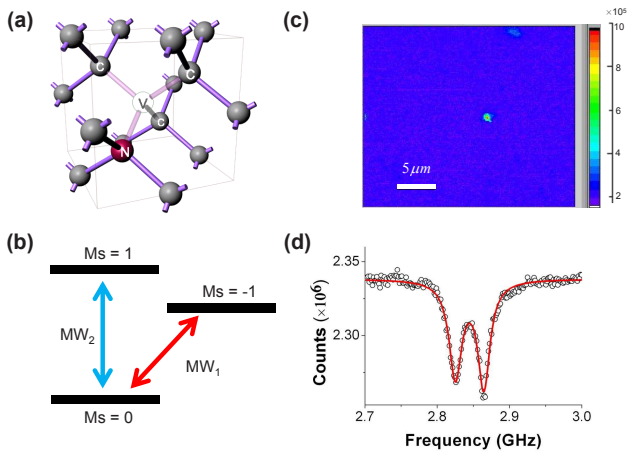}
    \caption{
    (a) Atomic structure of the NV center in diamond \cite{jele2}. The
    NV center comprises a substitutional nitrogen (N), and a neighboring
    vacancy (V).
    (b) Energy level diagram of the electronic ground state showing the
    zero field splitting. 
    (c) Fluorescence microscopy image of the single NV center.
    (d) Optically detected magnetic resonance spectrum for the single NV center
    obtained by a frequency sweep using 4000 averages. Lorentzian peaks (red line)
    were fitted to the experimental spectrum (black circles).}
    \label{fig1}
\end{figure}

We realize the RDJ algorithm by encoding both qubit and an auxiliary state in the $S=1$ electron spin of an NV center. To the best of our knowledge, this is the first room-temperature implementation of a quantum algorithm on individual spins. To carry out the single-qubit RDJ, we need Hadamard gates and f-controlled gates. The former produces the superposition of states from the input state and, after the evaluation function has run, reconverts the superposition to a detectable polarization output.
The f-controlled gate is defined as $V_{f}|z\rangle = (-1)^{f(z)}|z\rangle$, where $z=$0, 1, and $f(z)$ is embodied by four functions with $f_{1}(z)=0$ and $f_{2}(z)=1$ for constant
functions, and with $f_{3}(z)=z$ and $f_{4}(z)=1-z$ corresponding to balanced functions. As a result, for a two-level system, $V_{f_i}$ can be written explicitly as $V_{f_1}=-V_{f_2}= \left(\begin{smallmatrix} 1 & 0 \\ 0 & 1 \end{smallmatrix}\right)$ and $V_{f_3}=-V_{f_4}=\left(\begin{smallmatrix} 1 & 0 \\ 0 & -1 \end{smallmatrix}\right)$.
For the two levels $|0\rangle$ and $|1\rangle$, the qubit is initially prepared in $|0\rangle$. After a Hadamard gate, followed by the f-controlled gate $V_{f_i}$, the state of the system evolves to
$$\frac {1}{\sqrt{2}}[(-1)^{f_{j}(0)}|0\rangle + (-1)^{f_{j}(1)}|1\rangle],$$
with $j=$1, 2, 3, 4. So the constant (balanced) function, after the second Hadamard gate, will evolve to $(-1)^{f_{j}(0)}|0(1)\rangle$, which can be identified by a single readout. To simplify the experimental realization, we replace the Hadamard gate by a selective $\frac{\pi}{2}$ microwave pulse, which inverts the output of the refined DJ version compared to the original algorithm scheme but otherwise does not change our conclusions.

\begin{figure}[htbp]
\centering
\includegraphics[width=0.9 \columnwidth]{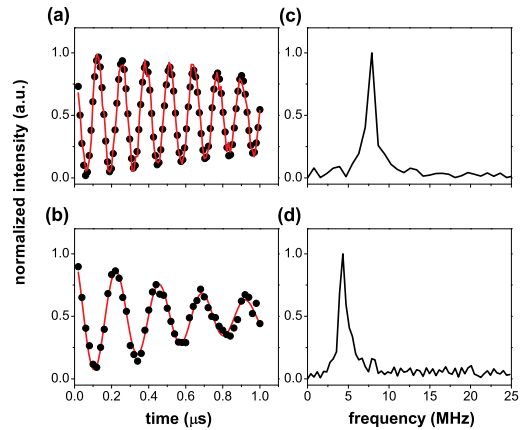}
    \caption{
    Transient nutation of the electron spin between ground
    state sublevels of the single NV center, where the upper plot is for
    nutation between $m_{s}=0$ and $m_{s}=-1$ sublevels and the lower
    one for the nutation between the $m_{s}=0$ and $m_{s}=1$
    sublevels. The experimental data (black solid
    circles) were fitted  by a damped sine function (red line)
    written as $y = y_0 + A e^{-R t} \cos(\omega t)$, with the
    intensity offset $y_0$ and the amplitude $A$ to be
    $0.5$. This normalization is independent from the decay rate $R$ and
    nutation frequency $\omega$. The fast Fourier transform (FFT)
    was applied to obtain the Rabi frequencies.
    (a) Nutation experiment at frequency $2.8254~$GHz ($MW_1$) with $1$ million
    averages. Its FFT spectrum (b) shows a $7.87~$MHz Rabi
    frequency. (c) Nutation experiment at frequency $2.8644~$GHz ($MW_2$)
    with $1$ million averages. Its FFT spectrum (d) shows a $4.26~$MHz Rabi
    frequency.}
    \label{fig2}
\end{figure}

The structure and the ground state of the NV center we employ are depicted in Fig. \ref{fig1}, where the defect includes a substitutional nitrogen atom and a vacancy in the nearest neighbor lattice position (Fig. \ref{fig1}(a)). It is negatively charged since the center comprises six electrons, two of which are unpaired. Our sample is a commercial diamond nanocrystal (nominal diameter 25 nm). Fig. \ref{fig1}(c) shows an image of the nanocrystal detected by the fluorescence microscopy.
Since we did not apply an external magnetic field, the Hamiltonian of the NV center is given by \cite{Hamiltonian_NV}:
\begin{equation}
\label{eq£º1}
H =  \hat{\textbf{S}}\overleftrightarrow{\textbf{D}}\hat{\textbf{S}}
   + \hat{\textbf{S}}\overleftrightarrow{\textbf{A}}\hat{\textbf{I}},
\end{equation}
where $\hat{\textbf{S}}$ and $\hat{\textbf{I}}$ are the spin operators associated with  electron and nucleus, respectively. The second term represents the hyperfine interaction between these spins, which is not employed in our operations but is the source of electron-spin dephasing. The optically detected magnetic resonance spectrum of the NV center (Fig. \ref{fig1}(d)) shows that the magnetic hyperfine interaction ($A\simeq 2$ MHz \cite{N splitting}) was not resolved in our experiments. The first term is the zero-field splitting (or fine structure) caused by mutual interaction of the two uncoupled electrons. This dipolar term can be written as $H_D = D[S^2_z - \frac{1}{3}S(S+1)] + E(S^2_x-S^2_y)$. For diamond nanocrystals, the value of $E$ is usually non-zero because of the strain induced by its vicinity to the surface, which assures that all degeneracies of the triplet ground state are lifted. We find $D=2.8449$ GHz and $E=19.5$ MHz (Fig. \ref{fig1}(b)), corresponding to a splitting of 2.8644 GHz between the states $|0\rangle$ ($m_{s}=0$) and $|1\rangle$ and 2.8254 GHz between $|0\rangle$ and $|-1\rangle$. In our experiment, we encoded the qubit in  $|0\rangle$ and $|-1\rangle$, and took the level $|1\rangle$ as an auxiliary state.

The experiments were carried out with a home-built confocal microscope operated at room temperature. The sample, mounted at the focus of the microscope, is illuminated by a
diode-pumped solid-state laser (Oxxius, SLIM-532S-50-COL-PP) at a wavelength of $\lambda = 532~$nm. A piezoelectric scanner (Physik Instrumente, P-562.3CD) was used to control the focus of an oil immersion objective (Olympus, PlanApoN, 60x, NA$=1.42$). The NV center fluorescence is separated from the excitation laser with a long wave pass filter (Semrock, BLP01-635R) and then collected by a silicon avalanche photodiode (APD) (Perkin Elmer, SPCM-AQRH-13). We constructed two synchronized microwave channels that provide the setup with the ability to output microwave pulses with two different frequencies. The microwave is coupled to the sample by a $20$ $\mu$m diameter copper wire acting as antenna. The whole system is orchestrated by a word generator (SpinCore Technologies, PBESR-PRO-350).

As a preparation for the RDJ algorithm, we have first accomplished coherent spin resonance between the ground state sublevels. Fig. \ref{fig2} shows the transient nutations of a single NV center. The initialization of state $|0\rangle$ with $>90\%$ probability is achieved by a $5$ $\mu s$ green excitation, followed by a waiting time of $5$ $\mu s$ \cite{Room-temperature coherent coupling of single spins in diamond}. A microwave pulse of variable duration was then applied to the NV center, and the spin state was read out by monitoring the fluorescence intensity. The experimental data show a periodic modulation of the fluorescence signals between the $|0\rangle$ and $|-1\rangle$ (Fig. \ref{fig2}(a)) and
between the $|0\rangle$ and $|1\rangle$ (Fig. \ref{fig2}(b)), respectively. From the figures, we can extract the Rabi frequencies under microwave irradiation ((Fig. \ref{fig2}(c) and (d)), which is important for performing the gates in the RDJ algorithm below. We used $64~$ ns and $118~$ ns long $\pi$ pulses for the two microwave channels, respectively. The decay of coherent oscillations is due to electron spin dephasing, affecting the visibility of the spin state in a predictable manner.

\begin{figure}[htbp]
\centering
\includegraphics[width=1.0 \columnwidth]{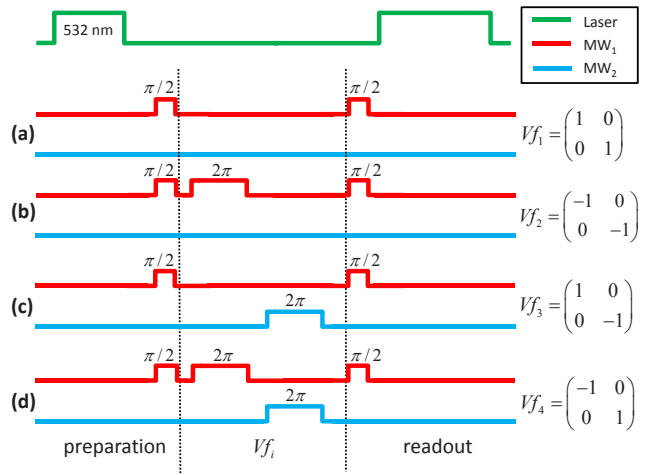}
    \caption{Diagram of the experimental pulse sequences used to realize the
    RDJ algorithm. The $532$ nm laser (green line) was
    used to initialize the state of the NV center to $|0\rangle$ and
    was shut off during the algorithm. Later, the laser was switched on
    again for detection. $MW_1$ (red line) and $MW_2$ (blue line) are two
    microwave channels which excite different transitions selectively.
    The first $MW_1$ $\frac{\pi}{2}$ pulse was used to generate a
    superposition state in the qubit and the last $MW_1$ $\frac{\pi}{2}$
    pulse was taken for detection by reconverting coherence to population
    differences. The $V_{fi}$ operations ($i = 1, 2, 3, 4$) were realized
    by combinations of (both $MW_{1}$ and $MW_{2}$) $2\pi$ pulses, as shown
    between the vertical dashed lines.}
    \label{fig3}
\end{figure}

In the implementation of RDJ, after preparing the initial state $|0\rangle$, we applied a selective $\frac{\pi}{2}$ microwave pulse in the $MW_1$ channel, yielding the superposition. The following f-controlled gate operations have been implemented by combinations of $2\pi$ pulses in the four possible cases (Fig. \ref{fig3}). Making use of the auxiliary state $|1\rangle$, we applied a $2\pi$ pulse in the $MW_2$ channel, introducing a $\pi$ phase shift on the state $|0\rangle$. This is equivalent to a $\pi$ rotation about $Z$ axis in the subspace spanned by $|0\rangle$ and $|-1\rangle$ \cite{2pi}. Switching the 532 nm laser on again, the final result was read out from the fluorescence collected by the APD. Fig. \ref{fig3} shows the four groups of microwave pulse combinations we used, corresponding to the four f-controlled gate operations. The intervals between pulses are set to zero in our experiments and we merged some pulses into a longer one in order to simplify the pulse sequence. For example, a $MW_1$ $~\pi$ pulse was used in Fig.\ref{fig3}(a) instead of the two
$\frac{\pi}{2}$ pulses. This increases the selectivity of these pulses, but the pulses are still broad enough to cover the resonance line.

\begin{figure}[htbp]
\centering
\includegraphics[width=0.9\columnwidth]{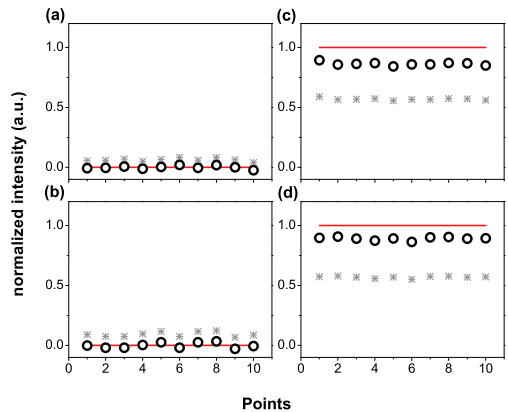}
    \caption{
    The output of the RDJ algorithm is illustrated by the
    intensity of the fluorescence, where (a) and (b) with the weakest intensity
    correspond to the constant function, and (c) and (d) with the strongest
    fluorescence indicate the balanced functions. While the original data
    (gray stars) suffer from dephasing during the pulse implementation,
    the outputs are clear enough to demonstrate the differences of f-control functions.
    The black circles are plotted with the dephasing effect in spin nutation compensated.}
\label{fig4}
\end{figure}

The results of the RDJ algorithm are shown in Fig.\ref{fig4}. Each point (gray stars) represents an individual experiment with $50~$ million averages, and we did four sets of experiments corresponding to the four f-controlled gate operations. Note that since we have used selective $\frac{\pi}{2}$ pulses in place of the Hadamard gates, the output is inverted with respect to the original algorithm. The weakest fluorescence intensity (minimum population in $|0\rangle$) thus corresponds to the constant functions $V_{f1}$ and $V_{f2}$, and the strongest intensity (maximum population in $|0\rangle$) indicates balanced functions $V_{f3}$ and $V_{f4}$. The fluorescence difference obtained for constant and balanced functions is $56.9\%$ (gray stars in Fig. \ref{fig4}), which is reduced mostly due to dephasing during the pulse operations (estimated as $\sim59.6\%$ from Fig. \ref{fig2}). Nevertheless, it is clear enough to illustrate the accomplishment of the RDJ. For clarity, we have compensated the dephasing effect using the results in Fig. \ref{fig2}. As a result, a small deviation of the compensated experimental data (black circles in Fig. \ref{fig4}) from the theoretical expectation (red lines in Fig. \ref{fig4}) remains, due to operational imperfections of the microwave pulses.

In contrast to room-temperature experiments with nuclear magnetic resonance using spin ensembles \cite {nmr}, our experiment works on a solid-state quantum system. As a result, we have achieved a pure-state QC implementation at room temperature. With respect to other systems \cite {qd1,qd2,ion} for coherently manipulating individual spins, our implementation without cryogenic requirements greatly reduces the experimental challenge for carrying
out QC. Moreover, as our qubit in the NV center can be fixed and manipulated exactly, the QC operation in our case is deterministic and efficient.

The full demonstration of the power of quantum algorithms requires large-scale QC. Optical coupling of spatially separate NV centers might be achieved by putting the centers in optical cavities, which enhances both the zero phonon line and the collection efficiency of the emitted photons. Considerable efforts have been made for fabricating thin, single-crystal diamond membranes \cite{mem}, whispering-gallery mode resonators \cite {wgm} and photonic
band-gap microcavities \cite {cavity}. However, since these systems usually work well only at low temperatures, and since the excited states of the NV centers are not well protected from decoherence at room temperature due to spin-orbit coupling \cite {jele09}, cryogenic
operation seems necessary for extending the NV-center QC architecture in this way. Therefore, how to accomplish a large-scale QC at room temperature is still an open question \cite {nv}. Moreover, once more qubits are involved in the system, the required operations get more complicated and time-consuming. This implies that we need to fasten the operations or to effectively suppress decoherence. Some first explorations into these aspects have been
reported \cite {giga, bath}. Nevertheless, our present experiment has clearly shown the unique opportunity provided by NV centers to study the physics and application of single spins, and also demonstrated the great potential of the NV system for QC.

In summary, we have accomplished a RDJ quantum algorithm using only the electron-spin of a single NV center at room temeprature by exploiting the $S=1$ character of this system. Our experimental data (after compensation of a systematic dephasing effect) fit the theoretical prediction well with a small deviation due to pulse imperfections. Although building a scalable quantum computer is pretty hard with current technology, successful implementations of existing quantum algorithms using available QC building blocks would be definitely helpful in stimulating inventive new ideas and further technologies. In this sense, it is of great importance for our experiment to demonstrate the power of QC at room temperature in a real solid system, using minimal quantum resources. The possibility of carrying out quantum superposition and interference at room temperature makes future work toward large-scale room-temperature QC architectures worthwhile.

F. Shi acknowledges W. Gao for helpful discussion. This work was supported by the NNSFC, the CAS, the Ministry of Education, PRC, and the 973 program (contract no. 2007CB925200). The German side was supported by the Volkswagen Stiftung through the program "Integration of molecular components in functional macroscopic systems" and by the Bundesministerium f{\"u}r Bildung und Forschung (contract no. 03N8709).

\end{document}